\documentclass[10pt,a4paper]{iopart}
\usepackage{iopams}  
\usepackage{graphicx,psfrag,bbm,latexsym,color,dcolumn,bm,dsfont,bbm,color,
mathrsfs,bbold,latexsym,amsfonts,amssymb}
\usepackage{epsfig}

\def\defi{{\buildrel \;def\; \over =}}
\newcommand{\be}{\begin{equation}}
\newcommand{\ee}{\end{equation}}

\newcommand{\mediaT}[1]{\left\langle #1 \right\rangle}

\begin{document}

\title{Small-world of communities:\\ communication and correlation of the meta-network}

\author{Massimo Ostilli$^{1,2}$, J. F. F. Mendes$^{1}$}
\address{$^1$\ Departamento de F{\'\i}sica da Universidade de Aveiro, 3810-193 Aveiro, Portugal.}
\address{$^2$\ Center for Statistical Mechanics and Complexity, Istituto 
Nazionale per la Fisica della Materia, 
Unit\`a di Roma 1, Roma 00185, Italy.}

\ead{massimo.ostilli@roma1.infn.it}

\date{\today}


\begin{abstract}
Given a network and a partition in $n$ communities, 
we address the issues ``how communities influence each other''
and ``when two given communities do communicate''.
We prove that, for a small-world network, among communities, a simple superposition principle applies 
and each community plays the role of a microscopic spin governed by a sort of effective TAP  
(Thouless, Anderson and Palmer) equations.
The relative susceptibilities derived from these equations calculated at finite or zero
temperature (where the method provides an effective percolation theory) give us 
the answers to the above issues.
As for the already studied case $n=1$, these equations  
are exact in the paramagnetic regions (at $T=0$ this means below the percolation threshold) and provide 
effective approximations in the other regions. 
However, unlike the case $n=1$,
asymmetries among the communities may lead, via the TAP-like structure of the equations,    
to many metastable states whose number, 
in the case of negative short-cuts among the communities,
may grow exponentially fast with $n$ and glassy scenarios with a remarkable presence of abrupt jumps take place.
Furthermore, as a byproduct, from the relative susceptibilities
a natural and efficient method to detect the community structure of a generic network emerges.
\end{abstract}


\maketitle

\section{Introduction} Recently, in the network science, the issue to find
the ``optimal'' community structure that should be present in a given random graph (a network)
$(\mathcal{L},\Gamma)$, $\mathcal{L}$ and $\Gamma$ being the set of the vertices
and bonds, respectively, has received much attention. 
The general idea behind the concept of community structure comes from the observation
that in many situations real data show an intrinsic partition of the vertices of the graph into $n$ groups, called
communities, $(\mathcal{L}=\cup_{l=1}^{n}\mathcal{L}^{(l)},\Gamma=\cup_{l\leq k=1}^{n}\Gamma^{(l,k)})$, 
such that between any two communities
there is a number of bonds that is relatively small if compared with the number of bonds present in each community. 
The partition(s) can be used to build a higher-level meta-network where the meta-nodes are now the communities  
(cells, proteins, groups of people, $\ldots$) and play
important roles in unveiling the functional organization inside the network.
In order to detect the community structure of a given network, many methods have been proposed 
and special progresses have been made by mapping the problem for identifying community structures to
optimization problems 
\cite{Blatt,Girvan}, by looking for $k-$clique sub-graphs \cite{Palla},
or by looking for clustering desynchronization \cite{Boccaletti}.
In general there is not a unique criterion. However, given 
a structure in communities, whatsoever be the method used, 
and assuming that the found partition $(\cup_{l=1}^{n}\mathcal{L}^{(l)},\cup_{l\leq k=1}^{n}\Gamma^{(l,k)})$ 
represents sufficiently well the intrinsic community structure of the given network \cite{Newmanrob}, 
there is still left
the fundamental issue about the true relationships among these communities.
Under which conditions, 
and how much two given communities, also in the presence of other communities, exchange information, 
how they influence each other, positively or negatively,
what is the typical state of a single community, what is the expected behavior for $n$ large, etc... are all issues
that cannot be addressed by simply using the above methods to detect the community structure.
In fact, all these methods, with the exception of Refs. \cite{Blatt} and \cite{Boccaletti}, 
are essentially based only on a topological (and, in most cases, local) analysis 
of the network. 
To uncover the real communication among the communities
we have to pose over the graph $(\mathcal{L},\Gamma)$ a minimal model in which the vertices
assume at least two states, and analyze their correlations. 
Confining the problem to the equilibrium case 
we have hence to use the Gibbs-Boltzmann statistical mechanics. 
Given a community structure of size $N$, if each node is associated with a spin $\sigma_i$, $i=1,\ldots,N$,  
then each meta-node is associated with the meta-spin  
$s^{(l)}=\sum_{i\in\mathcal{L}^{(l)}}\sigma_i/\sum_{i\in\mathcal{L}^{(l)}}1$, $l=1,\ldots,n$,
where the sums run only over the nodes of the $l$-th community. 
\begin{figure}[bh]
\epsfxsize=45mm \centerline{\epsffile{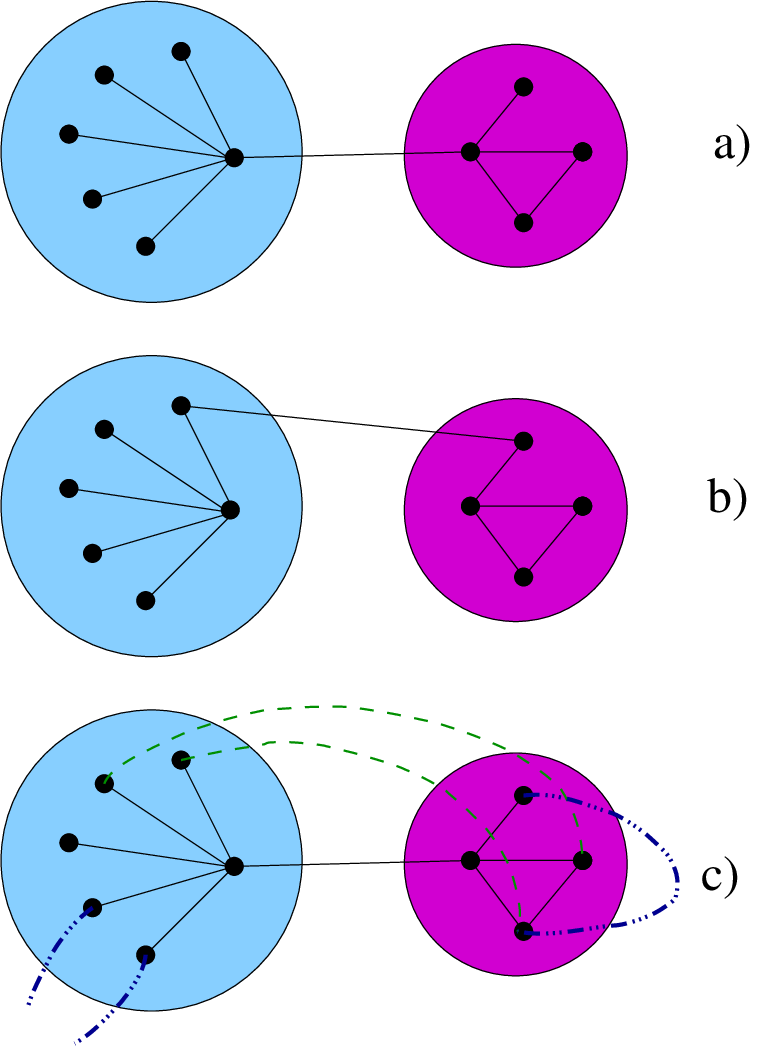}}
\caption{(Color on-line) An example with $n=2$, $|\mathcal{L}_0^{(1)}|=6$, $|\mathcal{L}_0^{(2)}|=4$, $|\Gamma_0^{(1,1)}|=5$,
$|\Gamma_0^{(2,2)}|=4$, $|\Gamma_0^{(1,2)}|=1$. Continuous lines represent short-range couplings $J_0^{(l,k)}$ 
of the pure model, whereas dashed and dot-dashed lines represent the additional random couplings $J_{i,j}^{(l,k)}$,
with $i\in\mathcal{L}_0^{(l)}$ and $j\in\mathcal{L}_0^{(k)}$, and are present only in the random model (panel c)). 
Panel a) and b) differ for the position 
of one single coupling $J_0^{(1,2)}$, nevertheless the example in a), compared with the case in b), 
at least at finite temperature (where the lengths of paths are important), benefits
clearly of a better communication (higher betweenness).} 
\label{fig1}
\end{figure}
We show then that the 
average magnetizations $m^{(l)}$ of the meta-spins 
obey special effective equations as if they were microscopic spins immersed in a ferro or glassy material.    
From these equations it will be then simple to derive the relative susceptibilities $\chi^{(l,k)}$, $l,k=1,\ldots,n$
among the communities revealing in an unambiguous way the communication and correlation properties of the
meta-network. 

\section{A minimal model}
If a Gibbs-Boltzmann distribution, $\exp(-\beta H)$, 
has been assumed, $H$ being some effective Hamiltonian, one can obtain the (adimensional) couplings $\beta J_{i,j}^{(l,k)}$ 
from the data of the given graph by isolating
the two vertices $i,j$ from all the others, and by measuring the correlation function
of the obtained isolated dimer $\mediaT{\sigma_i\sigma_j}'$, where $\mediaT{\cdot}'$ stands for the Gibbs-Boltzmann
average of the isolated dimer. 
The general problem is actually more complicated due to the presence of two sources of disorder since
both the set of bonds $\Gamma$, and the single couplings $\{J_{i,j}^{(l,k)}\}$, may change with time.
Assuming that the time scale over which this changes take place is much larger than that 
of the thermal vibrations of the spins, we have then to consider a disordered Ising model 
with quenched disorder. 
Here we specialize this general problem to the case of Poissonian disorder of the
graph, while we leave the disorder of the couplings arbitrary. We formulate the problem in
terms of Ising models on generic small-world graphs \cite{Watts}: given an arbitrary graph $(\mathcal{L}_0,\Gamma_0)$, 
\textit{the pure graph}, and 
an associated community structure $(\cup_{l=1}^{n}\mathcal{L}_0^{(l)},\cup_{l\leq k=1}^{n}\Gamma_0^{(l,k)})$,
in which each community has an arbitrary size, we consider a generic Ising Hamiltonian $H_0$ defined 
on this non random structure, \textit{the pure model}, characterized by arbitrary couplings $J_0^{(l,k)}$,
and, between any two sites $i,j$, 
with $i\in\mathcal{L}^{(l)}$ and $j\in\mathcal{L}^{(k)}$, 
we add some random connections (short-cuts) 
having average connectivities $c^{(l,k)}$ (more precisely, if $|\mathcal{L}_0^{(k)}|$ is the size of the $k$-th community, 
we introduce the directed random variables $c_{i,j}^{(l,k)}$ taking
the values $c_{i,j}^{(l,k)}=0,1$ with probabilities $1-c^{(l,k)}/|\mathcal{L}_0^{(k)}|$ and $c^{(l,k)}/|\mathcal{L}_0^{(k)}|$, respectively). 
Then, along with these random connections, random couplings $J^{(l,k)}$ with a quenched disorder are imposed,
and we study the corresponding random Ising model, \textit{the random model}, having Hamiltonian $H$ (see Fig. 1).
In a compact way $H$ is therefore given by
\begin{eqnarray}
\label{THEO00}
H\left(\{\sigma_i\}_{i=1}^{N}\right)=&H_0&\left(\{\sigma_i\}_{i=1}^{N}\right)+\Delta H\left(\{\sigma_i\}_{i=1}^{N}\right)
-\sum_{l=1}^n h^{(l)}\sum_{i\in\mathcal{L}_0^{(l)}}\sigma_i.
\end{eqnarray}
In Eq. (\ref{THEO00}) $H_0$ is the non random part of the Hamiltonian having non random 
couplings $J^{(l,k)}_{0}$ (typically short-range couplings but not necessarily) whereas $\Delta H$ is the random part of the
Hamiltonian involving only the long-range couplings $c^{(l,k)}_{i,j}J^{(l,k)}_{i,j}$,
finally $h^{(l)}$ is an arbitrary external field acting only on the $l$-th community.

\section{Equations of the meta-network}
In \cite{SW} we established a new general method to analyze critical phenomena on the small-world models
represented by Eq. (\ref{THEO00})
for the case in which there is one single community, $n=1$: given any arbitrary initial
graph $(\mathcal{L}_0,\Gamma_0)$, if we increase the average connectivity by $c$
through the addition of $Nc$ bonds randomly spread over the initial graph, 
under the only condition that be $c>0$,  
we have found an effective field theory that generalizes the Curie-Weiss mean-field theory via the equation  
\begin{eqnarray}
\label{THEO0}
m^{(\Sigma)}=m_0\left(\beta J_0^{(\Sigma)},\beta J^{(\Sigma)}m^{(\Sigma)}\right)
\end{eqnarray}
and that is able to take into account both
the infinite and finite dimensionality simultaneously present in small-world models. 
In Eq. (\ref{THEO0}) $m_0(\beta J_0;\beta h)$ represents the
magnetization of the pure model, \textit{i.e.}, without short-cuts, 
supposed known as a function of the short-range coupling $J_0$ and an arbitrary external field $h$,
whereas the symbol $\Sigma$ stands for the ferro-like solution, $\Sigma=F$, or the spin glass-like solution, $\Sigma=SG$,
and $J_0^{(\Sigma)}$ and $J^{(\Sigma)}$ are effective couplings.
Here we generalize this result to the present case of $n$ communities of arbitrary sizes and interactions; 
short-range and long-range (or short-cuts) couplings.
We show that, among the communities, a natural superposition principle applies and
we find that the $n$ order parameters, F or SG like, obey a system  
of equations which, a part from the absence of the Onsager's reaction term \cite{Onsager},
can be seen as an $n\times n$ effective system of TAP (Thouless, Anderson and Palmer) equations \cite{TAP} in which each community
plays the role of a single ``microscopic''-spin through its own order parameter, $m^{(\Sigma;l)}$, 
$l=1,\ldots,n$ \cite{LONG}.
In particular, in the simpler case in which there are no short-range couplings among different communities
($J_0^{(l,k)}=0$ for $l\neq k$) 
these self-consistent equations take the form 
\begin{eqnarray}
\label{THEO}
m^{(\Sigma;l)}&=&m_0^{(l)}\left(\beta J_0^{(\Sigma;l)}; \beta H^{(\Sigma;l)}+\beta h^{(l)}\right), \\
\beta H^{(\Sigma;l)}&=& \sum_{k=1}^n \beta J^{(\Sigma;l,k)}m^{(\Sigma;k)}, \nonumber 
\end{eqnarray}
where
$m_0^{(l)}(\beta J_0^{(l)};\beta h^{(l)})$ is the magnetization of the pure model for the $l$-th community, and
the effective couplings $J^{(\mathrm{F};l,k)}$, $J^{(\mathrm{SG};l,k)}$,
$J_0^{(\mathrm{F};l)}$ and $J_0^{(\mathrm{SG};l)}$ are given by
\begin{eqnarray}
\label{THEO2}
\beta J^{(\mathrm{F};l,k)}&\defi& c^{(l,k)}\int
d\mu^{(l,k)}(J)\tanh(\beta J),\nonumber \\ \beta
J^{(\mathrm{SG};l,k)}&\defi& c^{(l,k)}\int d\mu^{(l,k)}(J)\tanh^2(\beta
J),\nonumber \\ J_0^{(\mathrm{F};l)}\defi J_0^{(l)},&\quad& \beta
J_0^{(\mathrm{SG};l)}\defi \tanh^{-1}(\tanh^2(\beta
J_0^{(l)})). \nonumber
\end{eqnarray} 
In the above definitions $d\mu^{(l,k)}(J)$ 
stands for the probability distribution of the long-range coupling disorder between the $l$-th and $k$-th community.
It is easy to see that, if the size of the communities is parameterized as $|\mathcal{L}_0^{(l)}|=\alpha^{(l)}N$, 
with $\sum_{l=1}^n\alpha^{(l)}=1$, then
the connectivities must satisfy the balance equation
$\alpha^{(l)}c^{(l,k)}=\alpha^{(k)}c^{(k,l)}$,
so that, when $\alpha^{(l)}\neq \alpha^{(k)}$, the
effective couplings $J^{(\mathrm{F};l,k)}$ and $J^{(\mathrm{SG};l,k)}$ are not symmetric
even if the random couplings $J_{i,j}^{(l,k)}$ are symmetric. 
Though being not complex, the rigorous derivation of Eqs. (\ref{THEO}) is quite lengthy. 
We refer the interested reader to \cite{LONG}. However, if we make the natural assumption that the effective couplings 
satisfy a linear superposition principle, we then see that Eqs. (\ref{THEO})
are immediately derived from Eq. (\ref{THEO0}). 

As for one single community, Eqs. (\ref{THEO})
are exact in the paramagnetic region (P) 
while provide an effective approximation in the other regions.
More precisely, off the P region, for unfrustrated disorders Eqs. (\ref{THEO}) are exact 
up to $\mathop{O}(1/c^{(l,k)})$ terms and become exact also in the limit $c^{(l,k)}\to 0^+$, 
whereas for frustrated disorders Eqs. (\ref{THEO})
in general give only a qualitative effective description of the order parameters. 
We stress however that in both the cases the critical surfaces derived from our effective Eqs. (\ref{THEO}) are exact
(notice in contrast that the pure naive mean field equations give a wrong critical surface)
whenever, for any $l=1,\ldots,n$, it is $c^{(l,l)}>0$ or, if for some $l$, $c^{(l,l)}=0$, 
there is at least a chain of, say $h$, connected communities 
such that $c^{(l,l_1)}>0,c^{(l_1,l_2)}>0,\ldots,c^{(l_h,l_h)}>0$.
We remind the reader that the remarkable progresses reached in the context of mean field theory for disordered
models, as in the case of the ordinary TAP equations (or the Viana-Bray model \cite{Viana}), 
concern only models where short loops in the graph are absent, while in our approach the graph
$(\mathcal{L}_0,\Gamma_0)$ may be an arbitrary lattice, regular or not, 
and having loops of any length.   
The possibility to improve the theory off the P region is a formidable task
to be investigated in the future, however, as we show below, it should be already
clear the very interesting phase-transition scenario emerging from Eq. (\ref{THEO}).

In \cite{SW} ($n=1$) we established the general scenario of the critical behavior
coming from Eq. (\ref{THEO0}), stressing the differences between the cases $J_0\geq 0$ and $J_0<0$,
the former being able to give only second-order phase transitions with classical critical exponents, 
whereas the latter being able to give rise, for a sufficiently large connectivity $c$, to 
multicritical points in principle also with first-order phase transitions.  
The same scenario essentially takes place also for $n\geq 2$ provided that the $J_0$'s and the $J$'s be 
almost the same for all the communities, 
otherwise many other situations are possible. In particular, unlike the case $n=1$, relative antiferromagnetism
between two communities $l$ and $k$ is possible as soon as the $J_{i,j}^{(l,k)}$ have negative averages,
while internal antiferromagnetism inside a single community, say the $l$-th one, due to the presence of
negative couplings $J_0^{(l)}<0$, is never possible as soon as disorder is present \cite{Herrero2}. 
Less intuitive and quite interestingly, if we try to connect randomly with some added connectivity $c^{(l,k)}$
the $l$-th community having inside only positive couplings (``good'') to the $k$-th community having
inside only negative couplings (``bad''), not only the bad community gains
a non zero order, but even the already good community gets an improved order. In Fig. 2 we report an example.
\begin{figure}
\epsfxsize=80mm \centerline{\epsffile{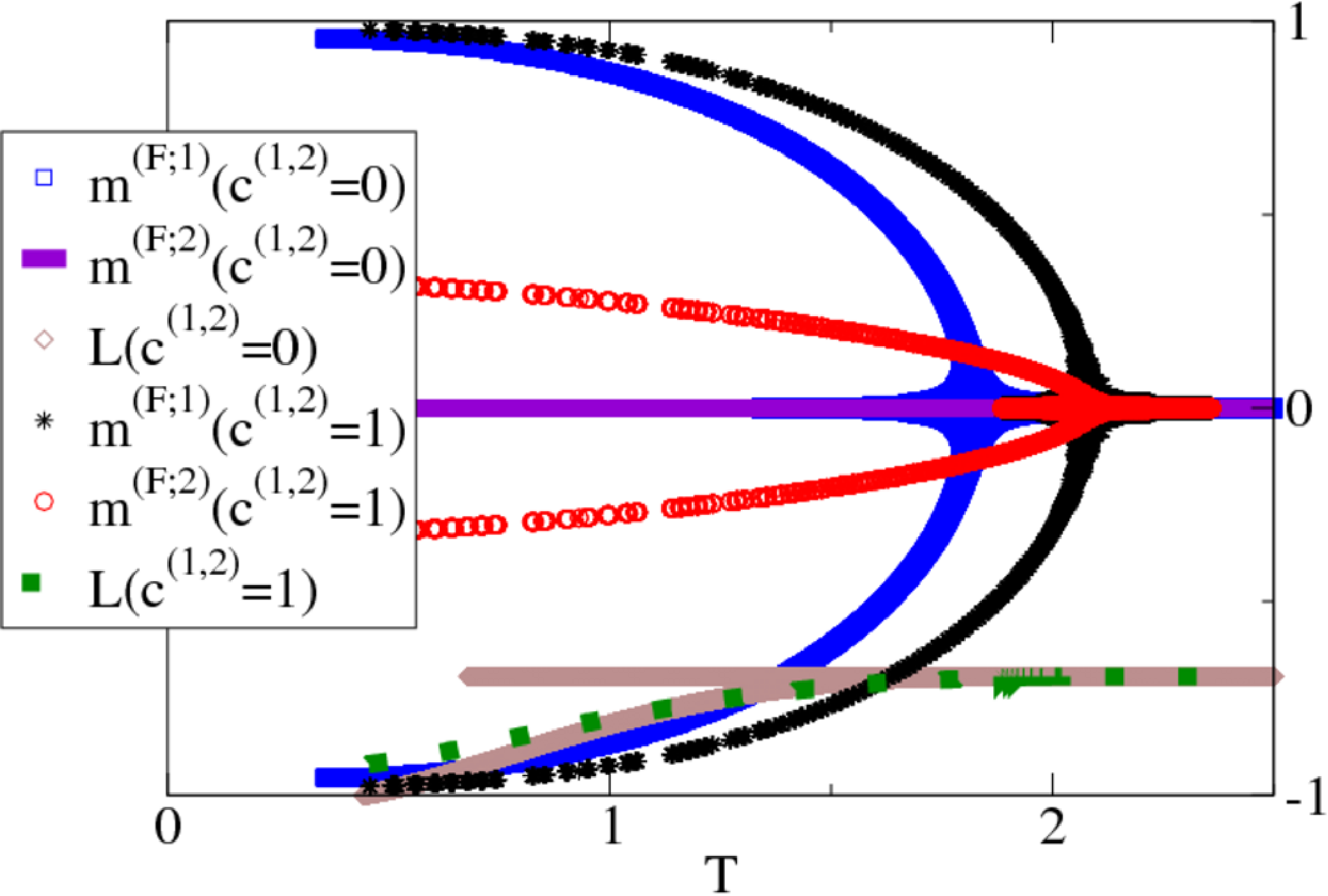}}
\caption{(Color online) Solutions of Eqs. (\ref{THEO}) with $n=2$
in the case $J_0^{(l,k)}\equiv 0$ (a generalized Viana-Bray model) with long-range couplings
$J^{(1,1)}=1$, $J^{(2,2)}=-1$ and $J^{(1,2)}=1$ ($J^{(1,2)}=-1$ also leads to the same plot),
and connectivities $c^{(1,1)}=c^{(2,2)}=2$, and $c^{(1,2)}=0$ or $c^{(1,2)}=1$.  
$L$ stands for the free energy term associated to each solution.} 
\label{fig2}
\end{figure}
However, with respect to the case $n=1$, another peculiar feature to take
into account is the presence of many metastable states. In fact, this is a general 
mechanism of the TAP-like structure of the equations: as we consider systems with an increasing number of communities,
the number of metastable states grows with $n$ and may grow exponentially fast in the case
of negative short-cuts. A metastable state can be made virtually stable (or, more precisely leading)
by forcing the system with appropriate initial conditions, by fast cooling, or by means of suitable external fields.
As a result, with respect to variations of the several parameters of the model 
(couplings, connectivities, sizes of the communities), 
the presence of metastable states may lead itself to first-order phase transitions
even when the $J_0$'s are all non negative. 
This general mechanism has been already
studied in the simplest version of these models, namely the $n=2$ Curie-Weiss model 
($J_0^{(l,k)}\equiv 0$ and $c^{(l,k)}\to\infty$),
where a first-order phase transition was observed to be tuned by the relative
sizes of the two communities and by the external fields \cite{Contucci}; 
moreover, first-order phase transitions have been observed
in simulations of a two dimensional small-world model with directed shortcuts \cite{Sanchez}.
In particular, in system of many communities,
$n\gg 1$, a remarkable and natural presence of first-order phase transitions (tuned by the several parameters) 
is expected which, if the $J$'s or the $J_0$'s are negative, 
reflects on the fact that the communities, at sufficiently low temperature, behave
as spins in an effective glassy state \cite{Parisi,Fisher}. Fig. 3 concerns the Curie-Weiss case with $n=3$;
as is evident, even for small $n$, the number of metastable states is rapidly growing.
\begin{figure}
\epsfxsize=80mm \centerline{\epsffile{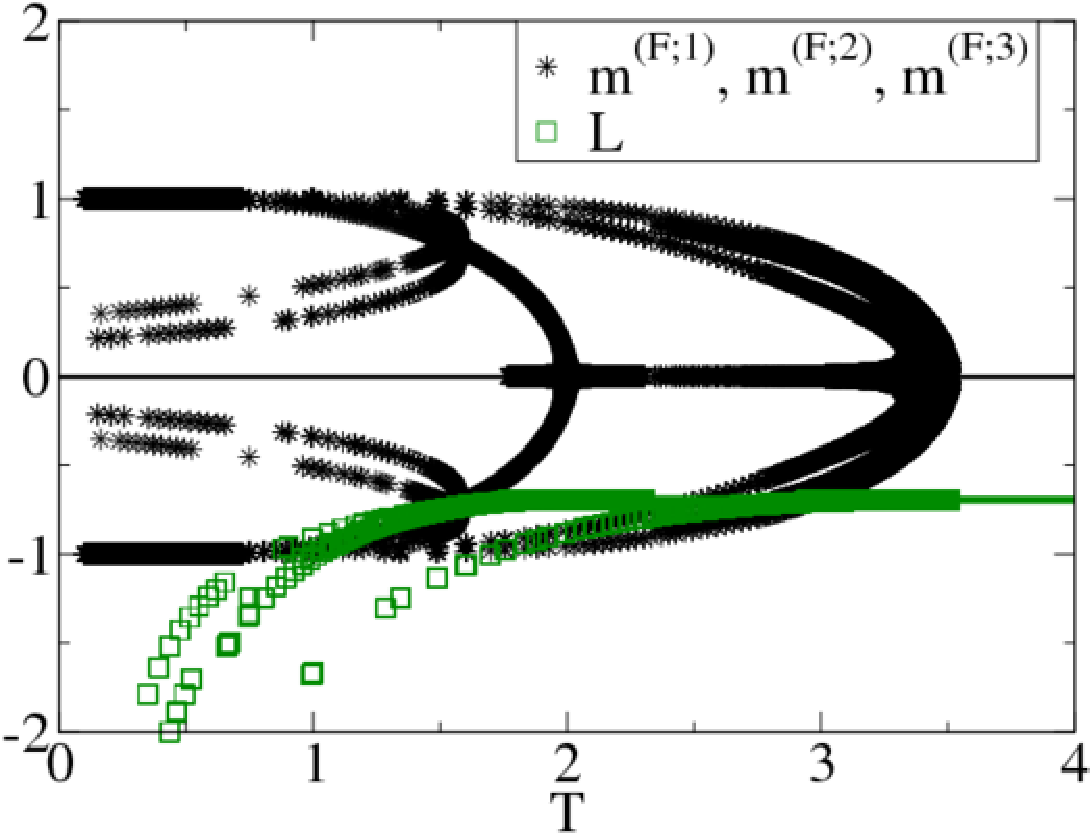}}
\caption{(Color online) Solutions of Eqs. (\ref{THEO}) 
with $n=3$ in the case $J_0^{(l,k)}\equiv 0$ and $c^{(l,k)}\to\infty$ (a generalized Curie-Weiss model), with
$J^{(l,l)}\equiv 3/c^{(l,l)}$ and $J^{(l,k)}\equiv -0.5/c^{(l,k)}$, for $l\neq k$.
We plot also the associated free energy term $L$.} 
\label{fig3}
\end{figure}

\section{Communication properties of the meta-network}
In general, how much two communities influence each other, is encoded in 
the matrix $\tilde{\chi}^{(l,k)}\defi\partial m^{(\Sigma;l)}/\partial (\beta h^{(k)})$,
the (adimensional) susceptibility of the random model which tells us how the $l$-th community reacts 
to a small variation occurred only (initially) in the $k$-th community. From Eq. (\ref{THEO}), or from its most general
form that includes also non zero short-range couplings among different communities, we have  
\begin{eqnarray}
\label{THEOsusc}
\bm{\tilde{\chi}^{(\Sigma)}}=\left(\bm{1}-\bm{\tilde{\chi}_0}\cdot\bm{\beta J^{(\Sigma)}}\right)^{-1}
\cdot \bm{\tilde{\chi}_0},
\end{eqnarray}
where we have introduced the matrix of the effective long-range couplings
$\beta J^{(\Sigma;l,k)}$, and 
$\tilde{\chi}_0^{(l,k)}$,
the adimensional susceptibility of the pure model. 
Note that in the case $J_0^{(l,k)}=0$ for $l\neq k$, $\bm{\tilde{\chi}_0}$ is a diagonal matrix 
whereas $\bm{\tilde{\chi}^{(\Sigma)}}$ is not. 
By looking for the points where $\bm{\tilde{\chi}^{(\Sigma)}}$ becomes singular,
from Eq. (\ref{THEOsusc}) we see immediately that the critical surface $\beta_c^{(\Sigma)}$ where the second-order 
transition takes place is solution of the following exact
equation 
\begin{eqnarray}
\label{THEOsusccrit}
\det\left(\bm{1}-\bm{\tilde{\chi}_0}\cdot\bm{\beta J^{(\Sigma)}}\right)=0.  
\end{eqnarray}

By sending $\beta\to\infty$, the theory can be projected in particular at zero temperature where, for positive couplings, a natural
effective percolation theory arises. Then, by using $\lim_{\beta\to+\infty} \bm{\beta J^{(\Sigma)}}=\bm{c}$, 
in the limit $\beta\to\infty$ Eq. (\ref{THEOsusc}) becomes
\begin{eqnarray}
\label{THEOsusc01}
\bm{\mathcal{E}}=\left(\bm{1}-\bm{\mathcal{E}_0}\cdot\bm{c}\right)^{-1}
\cdot \bm{\mathcal{E}_0},
\end{eqnarray}
where we have introduced
\begin{eqnarray}
\label{P3d}
\mathcal{E}_0^{(l,k)}\defi
\lim_{\beta\to+\infty} \tilde{\chi}_0^{(l,k)}\left(\left\{\beta J_0^{(l',k')}\right\};0\right).
\end{eqnarray}
Eq. (\ref{THEOsusc01}) tells us how $\bm{\mathcal{E}}$ changes as we vary $\bm{c}$,
being an exact equation as $\bm{c}$ belongs to the P region, \textit{i.e.}, below the percolation threshold $\bm{c}_c$,
solution of the following exact equation~\footnote{
In Eq. (\ref{P3d}) it is understood that we are considering only graphs
$(\mathcal{L}_0,\Gamma_0)$ for which the pure model has no critical temperature even for $T=0$.
In fact, if this is not the case, as occurs for instance if $(\mathcal{L}_0,\Gamma_0)$ is a $d_0$ dimensional lattice with
$d_0\geq 1$, the P region is shrunk to the single trivial point $c^{(l,k)}\equiv 0$ and we can
take effectively $\mathcal{E}_0^{(l,k)}=+\infty$ \cite{LONG}.
More precisely, if $\beta_{c0}<\infty$ or even if $\beta_{c0}=0$, as occurs in scale free networks
with exponent $\gamma\leq 3$ \cite{Review}, one should use Eqs. (\ref{P3d}) keeping $N$ finite. 
}
\begin{eqnarray}
\label{THEOsusccrit0}
\det\left(\bm{1}-\bm{\mathcal{E}_0}\cdot\bm{c}\right)=0. 
\end{eqnarray}

At $T=0$ there is no thermal dissipation and it is easier to analyze the communication properties. 
Given the arbitrary pure graph $(\mathcal{L}_0,\Gamma_0)$, 
and a community structure assignment which splits the set of the bonds $\Gamma_0$
in $n(n-1)/2$ sets, $\Gamma_0=\cup_{l,k}\Gamma_0^{(l,k)}$, in the pure graph the communities $l$ and $k$ communicate
if and only if $\mathcal{E}_0^{(l,k)}\neq 0$, whereas in the random graph if and only if 
$\mathcal{E}^{(l,k)}\neq 0$.
We can understand the communication process by observing that
the characteristic time $t_0^{(l,k)}$ to exchange a unit of information between the two communities $l$ and $k$
in the pure model grows as 
$t_0^{(l,k)}\propto\left(\mathcal{E}_0^{(l,k)}\right)^{-1}$
and, similarly, for the random model as
$t^{(l,k)}\propto\left(\mathcal{E}^{(l,k)}\right)^{-1}$.
From Eq. (\ref{THEOsusc01}) we see that 
in the pure model, if, for $l\neq k$, $\mathcal{E}^{(l,k)}_0=0$, 
the two communities $l$ and $k$ cannot communicate ($t^{(l,k)}_0\to\infty$),
but for any arbitrary small $c^{(l,k)}>0$ they communicate and the characteristic time decays with $\bm{c}$
approximately as (it is easy to see that for any $l$ is always $\mathcal{E}^{(l,l)}_0\geq 1$)~\footnote{
Analogous relations hold also at finite $T$, provided the source of the signal be sufficiently slow.}
\begin{eqnarray}
\label{Percotime}
t^{(l,k)}\propto\left(\mathcal{E}^{(l,l)}_0c^{(l,k)}\mathcal{E}^{(k,k)}_0+\mathop{O}\left(\bm{c}^2\right)\right)^{-1},
\end{eqnarray}
whereas, at higher order in $\bm{c}$, 
Eq. (\ref{THEOsusc01}) takes into account that the communities $l$ and $k$ can communicate also indirectly via chains
of other intermediate communities.
In general, if the pure graph has an own dimension $d_0$ (possibly fractal) sufficiently high, $d_0\geq 1$, one has 
$\mathcal{E}^{(l,k)}_0\to\infty$ for $N\to\infty$, so that we have $t_0^{(l,k)}\to 0$ and then also
$t^{(l,k)}\to 0$; \textit{i.e.}, the communities communicate instantaneously (they percolate). 
However, in the random model, even if $\mathcal{E}^{(l,k)}_0$ is finite,  
when $\bm{c}$ approaches the percolation threshold surface $\bm{c}_c$, then we have $t^{(l,k)}\to 0$, a peculiar feature which is possible
only in the random model.
We stress again that the graph $(\mathcal{L}_0,\Gamma_0)$ is completely arbitrary.
So, for example, for $n=1$, it is easy to check the consistency of our  
effective percolation theory (\textit{i.e.}, we recover the same critical surface) 
with the recent model introduced in \cite{Newman_Clustering} where the 
classical Erd$\mathrm{\ddot{o}}$s-R$\mathrm{\acute{e}}$nyi random graph \cite{Classical} is generalized to include a finite clustering.
However, whereas in \cite{Newman_Clustering} the percolation analysis is performed by 
\textit{ab initio} calculations starting directly from graph theory elements 
(at least this seems in principle possible for graphs having regular loops), 
our effective percolation theory requires to perform a simulated annealing toward $T=0$ of the non
random model defined over the pure graph $(\mathcal{L}_0,\Gamma_0)$ and immersed in a small external field.
At each small but finite $T$ we make the simulation, then, once the relevant
quantities like $\chi_0$ or $m_0$ are obtained for small $T$,
the percolation properties of the random graph $(\mathcal{L},\Gamma)$ are easily calculated.
Therefore,in our effective percolation theory, in Eq. (\ref{THEOsusc01}) the matrix $\bm{\mathcal{E}_0}$ represents
an input data. In general, it can be sampled efficiently by simple simulated annealing procedures by using Eq. (\ref{P3d})
since the problem is mapped to an unfrustrated Ising model ($\beta J_0^{(l,k)}\geq 0$). 

The matrix $\bm{\mathcal{E}_0}$ (or, more in general, at finite $T$ the matrix $\bm{\chi}$),
leads also to a natural new criterion to detect community structures:
given an hypothetical value $n$,
by a suitable generalization of the modularity introduced by Girvan and Newman \cite{Girvan} which makes use
of $\bm{\mathcal{E}_0}$ rather than that of the adjacency matrix, we can define a ``measure'' which takes into account
paths of arbitrary length rather than links, and the resulting community structure coincides with that partition
such that the communication among all the communities is minimal \cite{LONG}.
In the algorithm proposed in \cite{Girvan}, given the network, one removes
the link having the highest betweenness, where the concept of betweenness,
a measure of the centrality of the given link, 
can be given in several ways. In particular the betweenness of a link can be
defined as the number of geodesic paths (the shortest path connecting two sites) passing through it.
After deleting many times the links having the highest betweenness, 
a partition of the original network in communities can be obtained and
a measure of the quality of assignment in communities is given as
\begin{eqnarray}
\label{Perco7}
Q=Q_1=\sum_{l}\left[e_1^{(l,l)}-\left(a_1^{(l)}\right)^2\right],
\end{eqnarray}
where $e_1^{(l,k)}$ is  
the fraction of all bonds connecting the communities $l$ and $k$, 
and $a_1^{(l)}$ is defined as $a_1^{(l)}\defi\sum_{k}e_1^{(l,k)}$.
The term $(a_1^{(l)})^2$ in Eq. (\ref{Perco7})
represents the expected fraction of bonds falling inside the community $l$ when
their ends are connected randomly. 
Thanks to the presence of the term $(a_1^{(l)})^2$ in Eq. (\ref{Perco7}), 
$Q_1$ gives measure 0 when one considers
the trivial case in which $\Gamma_0$ is a single community ($n=1$), and partitions
that maximize $Q_1$ correspond to best community structures.
We can consider however other similar measures that takes into account not only bonds,
but also, for example, paths of two consecutive bonds. In general we can define 
\begin{eqnarray}
\label{Perco8}
Q_h=\sum_{l}\left[e_h^{(l,l)}-\left(a_h^{(l)}\right)^2\right],
\end{eqnarray}
where now $e_h^{(l,k)}$ is the fraction of all paths of length not greater than $h$
connecting the two communities $l$ and $k$, 
and $a_h^{(l)}\defi\sum_{k}e_h^{(l,k)}$.
Again we have that its square represents the expected fraction of paths of length not greater than
$h$ having both ends inside the community $l$ when they are connected randomly, 
and makes the measures (\ref{Perco8}) non trivial.
When $h\to\infty$, the matrix $\bm{e}$ is proportional to the matrix $\bm{\mathcal{E}_0}$.
It is important to note that, at $T=0$, the algorithm we propose to detect a community
structure coincides with that of Newman and Girvan for the case of geodesic betweenness
(this can be seen from the combinatorial meaning of the matrices $\mathcal{E}^{(l,k)}_0$ 
or $\mathcal{E}^{(l,k)}$ \cite{LONG}),
but the measure associated to the found partition is given by using $Q_\infty$, and not $Q_1$:
essentially $Q_1$ measures the relative lack of links among different communities, while
$Q_\infty$ measures the relative lack of communication among different communities.

We point out that, in the pure model, having some bonds between the $l$-th and $k$-th communities 
does not guarantee that the condition $\mathcal{E}^{(l,k)}_0>0$ be satisfied.
In fact, it is not difficult to see that to have $\mathcal{E}^{(l,k)}_0>0$ it is necessary that
the number of paths between the $l$-th and the $k$-th communities be at least of order $N$ \cite{LONG}. 
Note also that such a requirement does not exclude the possibility 
that even a single bond between the two communities be enough, provided that through this bond passes
at least $\mathop{O}(N)$ paths (high betweenness, or centrality; see Fig. 1).
It should be then clear that measures based on a local analysis, 
and on elementary use of the adjacency matrix,
cannot capture the real communication properties. 
We illustrate now in a simple example, analytically feasible in our approach, 
how remarkable can be the differences in taking into account
just links or, more properly, paths of any length.
Such differences become actually much more important and interesting at finite temperature,
where the length of each single path affects the susceptibility, while at $T=0$ the length
of a path does not play any role. However, for simplicity in the following we will
consider only the case $T=0$.
Suppose we have found $n$ Erd$\mathrm{\ddot{o}}$s-R$\mathrm{\acute{e}}$nyi sets, 
having intra averages connectivities $c^{(l,l)}$ and connected with each other 
randomly with inter averages connectivities $c^{(l,k)}$, $l\neq k$. 
In this case, we have $\bm{\chi_0}\equiv \bm{1}$, therefore, from Eqs. (\ref{THEOsusc01}) and (\ref{P3d}),
we get immediately $\bm{\mathcal{E}}=(\bm{1}-\bm{c})^{-1}$, which is very different
from the adjacency matrix $\bm{c}$. This can be well understood by observing that, 
below the percolation threshold surface given by Eq. (\ref{THEOsusccrit0}),
it holds $\bm{\mathcal{E}}=(\bm{1}-\bm{c})^{-1}=\bm{1}+\bm{c}+\bm{c}^2\ldots$, where each term in the sum takes into
account the presence of paths of length 0, 1, 2,$\ldots$ respectively.
Let us apply this result to the following six examples with $n=2$ and $n=3$ communities:
\begin{eqnarray}
\fl
\label{Ex1}
\bm{c}_1=\left(
\begin{array}{l}
  0.35 \quad 0.3 \\
  0.3 \quad 0.35
\end{array}
\right) \quad
\bm{c}_2=\left(
\begin{array}{l}
  0.4 \quad 0.3 \\
  0.3 \quad 0.3
\end{array}
\right) \quad
\bm{c}_3=\left(
\begin{array}{l}
  0.6 \quad 0.3 \\
  0.3 \quad 0.1
\end{array}
\right) \quad
\bm{c}_4=\left(
\begin{array}{l}
  0.7 \quad 0.3 \\
  0.3 \quad 0.0
\end{array}
\right) \quad
\nonumber
\end{eqnarray}
and 
\begin{eqnarray}
\label{Ex2}
\bm{c}_5=\left(
\begin{array}{l}
  0.4 \quad 0.1 \quad 0.1 \\
  0.1 \quad 0.2 \quad 0.1 \\
  0.1 \quad 0.1 \quad 0.1 \\
\end{array}
\right) \quad
\bm{c}_6=\left(
\begin{array}{l}
  0.6 \quad 0.1 \quad 0.1 \\
  0.1 \quad 0.1 \quad 0.1 \\
  0.1 \quad 0.1 \quad 0.0 \\
\end{array}
\right).
\nonumber
\end{eqnarray}
In all the above examples the matrix $\bm{c}$ is always normalized, so that we have always $\bm{e}=\bm{c}$.
Concerning the matrix $\bm{e}_\infty$, it coincides with the matrix $\bm{\mathcal{E}}=(\bm{1}-\bm{c})^{-1}$ 
to be normalized to 1.
The results corresponding to the above six examples, together with the modularities $Q$ and $Q_\infty$, and the maximum
eigenvalue $\lambda$ of the matrix $\bm{c}$, 
give, respectively:
\begin{eqnarray}\fl
\label{Ex3}
\bm{e}_{\infty;1}=\left(
\begin{array}{l}
  0.406 \quad 0.094 \\
  0.094 \quad 0.406
\end{array}
\right) \quad Q=0.2, \quad Q_\infty=0.312, \quad \lambda=0.650 
\nonumber
\end{eqnarray}
\begin{eqnarray}\fl
\label{Ex4}
\bm{e}_{\infty;2}=\left(
\begin{array}{l}
  0.437 \quad 0.094 \\
  0.094 \quad 0.375
\end{array}
\right) \quad Q=0.195, \quad Q_\infty=0.310, \quad \lambda=0.654
\nonumber
\end{eqnarray}
\begin{eqnarray}\fl
\label{Ex4}
\bm{e}_{\infty;3}=\left(
\begin{array}{l}
  0.562 \quad 0.094 \\
  0.094 \quad 0.250
\end{array}
\right) \quad Q=0.007, \quad Q_\infty=0.264,  \quad \lambda=0.740
\nonumber
\end{eqnarray}
\begin{eqnarray}\fl
\label{Ex5}
\bm{e}_{\infty;4}=\left(
\begin{array}{l}
  0.625 \quad 0.094 \\
  0.094 \quad 0.187
\end{array}
\right) \quad Q=-0.045, \quad Q_\infty=0.217,  \quad \lambda=0.810
\nonumber
\end{eqnarray}
and 
\begin{eqnarray}\fl
\label{Ex5}
\bm{e}_{\infty;5}=\left(
\begin{array}{l}
  0.333 \quad 0.023 \quad 0.042 \\
  0.023 \quad 0.249 \quad 0.032 \\
  0.042 \quad 0.032 \quad 0.221 \\
\end{array}
\right) \quad Q=0.320, \quad Q_\infty=0.463,  \quad \lambda=0.487
\nonumber
\end{eqnarray}
\begin{eqnarray}\fl
\label{Ex6}
\bm{e}_{\infty;6}=\left(
\begin{array}{l}
  0.436 \quad 0.027 \quad 0.049 \\
  0.027 \quad 0.191 \quad 0.025 \\
  0.049 \quad 0.025 \quad 0.172 \\
\end{array} 
\right) \quad Q=0.160, \quad Q_\infty=0.418, \quad \lambda=0.640.
\nonumber
\end{eqnarray}
From these examples we observe the following: \textit{i)} as expected $Q$ and $Q_\infty$
tend to be ``parallel'' measures, but they are quite far from being proportional;
\textit{ii)} from the point of view of communication, these six networks do not present very strong differences
(we expect stronger differences in other kinds of networks embedded in some geometry; 
as the case depicted in Fig. \ref{fig1}); 
\textit{iii)} as we pass from the case 1 to the case 4, and similarly from the case 5 to the case 6, \textit{i.e.},
toward more and more asymmetric cases, both $Q$ and $Q_\infty$ 
take lower and lower values, meaning that the partition has a poor meaning ($Q$ small),
or that the global communication among the communities is higher ($Q_\infty$ small).
This latter observation is less trivial and deserves attention: as a rule, asymmetric situations
benefit of a better communication. Consider in particular the cases 5 and 6 and compare
the matrix $\bm{c}$ (or $\bm{e}$) with the matrix $\bm{e}_\infty$: despite  
$e^{(1,2)}$ and $e^{(1,3)}$ be equal, we have that 
$e_\infty^{(1,3)}$ is almost twice $e_\infty^{(1,2)}$. This is due to the fact that,
between the second and third community, the latter benefits of a stronger asymmetry
with respect to the first community, which in turn has the largest density of intra bonds.
For the same reason, in the case 6, we note also that the matrix element $e_\infty^{(3,3)}$ is of the same order of
magnitude of $e_\infty^{(2,2)}$, despite being $e^{(3,3)}=0$.

\section{Conclusions}
%
Real-world networks present an intrinsic partition in
communities. 
However, despite important progresses, the absence of universality of the many proposed techniques, makes
the community detection an ``Art'' rather than a solid science \cite{Gulbahce}.
One weak point of these techniques lies on the fact  
that in most cases only the topology (\textit{i.e.}, the structure of the graph) - and often only a local topology - 
is taken into account and correlations never introduced. 
In particular, the real communication properties
among the communities cannot rely on a local analysis. On the other hand, starting from real-data
it is possible to define in an unambiguous way a minimal model, a disordered Ising model, able to take into account all the correlations, 
short- and long-range like, present on the given network. 
We then discover that, whatever be the given community structure, the exact relationship of the meta-nodes 
are regulated by a quite universal form of effective TAP equations which, through F- and SG-like order parameters 
and then the matrix $\bm{\chi}$ of the relative susceptibilities 
(in general completely different from the adjacency matrix), give rise to 
a reach variety of configuration and communication scenarios which are analytically and/or numerically feasible.
In particular, by simulated annealing procedures applied to the non random model ($\bm{c}=\bm{0}$), 
Eqs. (\ref{THEOsusc01})-(\ref{Percotime}) allow us to analyze
how the percolation and communication properties of the system changes 
when it becomes random ($\bm{c}\neq\bm{0}$). We can also use the same
technique to analyze the non random graphs itself for both studying the communication properties or its
community structure. 
We stress that there is no limitation in the choice of the non random graph $(\mathcal{L}_0,\Gamma_0)$
and in particular, as occurs in real-world networks, it can contain loops of any length
where therefore traditional mean-field theories developed for disordered systems (that suppose a tree-like
structure of the graph) cannot be applied. Our approach is exact in the P region (where however, we recall, correlations
of two spins neighbors of a same spin are not zero due to the presence of short loops) 
and on its boundaries (the critical surface).
The possibility to improve the method off the P region is under investigation.
We finally anticipate here that it is possible to generalize all the results to
the cases of small-world models with free-scale intra- and inter-connections among communities~\cite{NEW}.
 
\section{acknowledgments}
Work supported by the Socialnets project,
and PTDC/FIS/71551/2006.
M. O. thanks L. De Sanctis for useful discussions.\\


\begin{thebibliography}{10}




 

\bibitem{Blatt} M. Blatt, S. Wiseman, and E. Domany, Phys. Rev. Lett. \textbf{76}, 3251 (1996).

\bibitem{Girvan} M. E. J. Newman and M. Girvan, Phys. Rev. E \textbf{69}, 026113 (2004);
M. E. J. Newman, Phys. Rev. E \textbf {69}, 066133 (2004);
A. Clauset, M. E. J. Newman, and C. Moore, 
Phys. Rev. E \textbf {70}, 066111 (2004);
A. Capocci, V.D.P. Servedio, G. Caldarelli and F. Colaiori,
Physica A \textbf{352}, 669 (2005);
J. Reichardt and S. Bornholdt, Phys. Rev. E, \textbf{74}, 016110 (2006).





\bibitem{Palla} I. Derenyi, G. Palla, and T. Vicsek, Phys. Rev. Lett. \textbf{94}, 160202 (2005). 

\bibitem{Boccaletti} S. Boccaletti, M. Ivanchenko, V. Latora, A. Pluchino, and A. Rapisarda,
Phys. Rev. E \textbf{75} 045102(R) (2007). 


\bibitem{Newmanrob} B. Karrer, E. Levina, and M. E. J. Newman,  
Phys. Rev. E \textbf {77}, 046119 (2008).

\bibitem{Watts} D. J. Watts, S. H. Strogatz, Nature, \textbf{393}, 440 (1998).

\bibitem{SW} M. Ostilli and J. F. F. Mendes,
Phys. Rev. E \textbf{78}, 031102 (2008).

\bibitem{Onsager} L. Onsager, J. Am. Chem. Soc., \textbf{58}, 1486 (1936).

\bibitem{TAP} D. J. Thouless, P. W. Anderson and R. G. Palmer,
Phylos. Mag. \textbf{35}, 593 (1977).

\bibitem{LONG} M. Ostilli and J. F. F. Mendes, Phys. Rev. E \textbf{80}, 011142 (2009). 

\bibitem{Viana} L. Viana, A. J. Bray, J. Phys. C: \textit{Solid State Phys.}
\textbf{18}, 3037 (1985).

\bibitem{Herrero2} This has been already seen by simulation in C. P. Herrero, Phys. Rev. E, \textbf{77}, 041102 (2008).

\bibitem{Contucci} P. Contucci, I. Gallo, S. Ghirlanda, arXiv:0712.1119.

\bibitem{Sanchez} A. D. S\'anchez, J. M. L\'opez, M. A. Rodr\'iguez, 
Phys. Rev. Lett., \textbf{88}, 048701-1 (2002).

\bibitem{Parisi} M. Mezard, G. Parisi, M.A. Virasoro,  
\textit{Spin Glass Theory and Beyond} (Singapore: World Scientific) (1987).

\bibitem{Fisher} K.H. Fischer and J.A. Hertz, \textit{Spin Glasses}, Cambridge University Press (1991).

\bibitem{Review} S.N. Dorogovtsev, A.V. Goltsev, J.F.F. Mendes,	
Rev. Mod. Phys. \textbf{80}, 1275 (2008).

\bibitem{Newman_Clustering} M. E. J. Newman, arXiv:0903.4009 (2009).

\bibitem{Classical} P. Erd$\mathrm{\ddot{o}}$s,
A. R$\mathrm{\acute{e}}$nyi, Publ. Math. Debrecen \textbf{6} 290 (1959).

\bibitem{Gulbahce} N. Gulbahce and S. Lehmann, BioEssays 30:934-938 (2008).

\bibitem{NEW} M. Ostilli and J. F. F. Mendes, to appear elsewhere. 




 

































%

























  



























\end{thebibliography}
\end{document}